\documentclass[doublespacing]{elsart}

\usepackage{epsfig}

\usepackage{amssymb}

\begin{document}

\begin{frontmatter}


\thanks[label2]{Corresponding author}

\title{Tagging single muons and other long-flying relativistic 
charged particles by ultra-fast timing in air Cherenkov telescopes}


\author[Munich]{R. Mirzoyan\thanksref{label2}},
\author[Lodz]{D. Sobczynska},
\author[Munich]{E. Lorenz},
\author[Munich]{M. Teshima}
\address[Munich]{Max-Planck-Institut f\"ur Physik, F\"ohringer Ring 6, 
D-80805 M\"unchen, Germany}
\address[Lodz]{University of \L \'od\'z, PL-90236 Lodz, Poland}
\begin{abstract}
Atmospheric air Cherenkov telescopes are successfully used for ground-based,
very high-energy (VHE) $\gamma$ ray astronomy. 
Triggers from the so-called single 
muon and other long-flying relativistic charged particle events are 
an unwanted 
background for the Cherenkov telescope. 
Because of low rate at $\sim TeV$ energies 
the muon background is unimportant. It is much more intense for telescopes 
with high photon sensitivity and low energy threshold. Below a few hundred 
GeV energy, the so-called muon background becomes so intense, that it can 
deteriorate the sensitivity of telescopes (the so-called 
"{\it muon-wall}" problem). From general considerations 
it can be anticipated that the signature of these particles should 
be a light pulse 
with a narrow time structure. In fact, simulations show that the pulses from 
muons have a very narrow time profile that is well below the time resolutions 
of nearly all currently operating telescopes. In this report we elaborate 
on the time profile of Cherenkov light from the so-called single muons and 
show that a telescope with ultra-fast time response can open a new dimension 
allowing one to tag and to reject those events.
\end{abstract}
\end{frontmatter}
%
%
%
%

%
%
%
\section{Introduction}
The window of ground-based $\gamma$ astronomy was opened in 
1989 by the observation 
of a strong signal from the first TeV source, the Crab Nebula, by the 
Whipple collaboration \cite{whipple}. 
A major breakthrough in the technique was the 
image parameterization suggested by Hillas in 1985 \cite{hillas}. This 
parameterization allowed the efficient separation 
of rare $\gamma$ ray events 
from the orders of magnitude more intense background from the charged 
cosmic rays (CR). Since then, this new field of astronomy 
has progressed very rapidly and all new source discoveries have been 
made by means of this new type of telescopes, the so-called imaging air 
Cherenkov telescopes (IACT). Currently, a new generation of very large 
IACTs \cite{VERITAS}, \cite{mori}, \cite{mariotti}, 
\cite{hofmann} 
have either started to operate or are in the final 
phase of completion. It is hoped that the energy window from a few 
tens of $GeV$ ($(30-200) GeV$, depending on the instrument) up to 
the multi-$TeV$, can thus be exploited. 
Already discussions and planning are underway
for future IACTs 
with a threshold close to $(5-10) GeV$. In order to achieve a very 
low energy threshold, it is necessary to build telescopes with a very large 
reflector area. With lower energy, the background caused by triggers
from hadron
showers decreases progressively, while other 
backgrounds become more dominant. One such backgrounds, namely 
air showers induced by cosmic electrons, cannot be strongly 
suppressed with current tools (except by using a telescope's 
angular resolution), 
because the electrons produce electromagnetic showers, 
in practice, indistinguishable from those of the $\gamma$. 
Below $(10-30) GeV$, 
depending on telescope location and observation direction, the earth's 
magnetic field deflects cosmic electrons out of the detection volume 
in the atmosphere. 
Another very likely irreducible background is caused 
by hadron interactions transferring a large fraction of their energy to a 
$\pi^{\circ}$, which subsequently decays into 2$\gamma$ and thus 
generates a dominantly electromagnetic shower. 
One of the main backgrounds is caused by Cherenkov light flashes from
single, long-flying relativistic 
charged particles high up in the atmosphere. This background is called, in 
the community, the muon background, but there can also be contributions from 
the muon parents, relativistic charged pions, kaons or protons. 
There may also be contributions from a fraction of single, straight,
long-flying electrons that can survive a few radiation lengths without
bremsstrahlung losses.
For clarifying the question about the particle types producing 
triggers in a Cherenkov
telescope
we performed special Monte Carlo simulations of 
hadron showers. 
In these simulations we followed cases in which almost the entire light 
came from either a) muons or from 
b) pions, kaons and protons. 
It turned out 
that the channel a) is the dominant one and that it is, at least 
by an order of magnitude, more frequent compared to channel b). For 
convenience further in the text we will use the name muon ($\mu$) background 
for all the above-discussed cases.  
The $\mu$ ($\pi$) can have a large angular deviation from their parent 
hadron shower's axis and thus can appear unaccompanied in the relatively 
small field of view of a Cherenkov telescope. That shall be one of the 
reasons
why they are 
called "single $\mu$ events". 
It is interesting to note that a wide field of view telescope 
can be helpful in rejecting the $\mu$ (hadronic) background: 
often together 
with the $\mu$ it will measure also
their parent particles (or showers).    
\section{The image shapes produced by muons}
An air Cherenkov telescope, depending on its photon sensitivity (light 
collection area and efficiency), can trigger on $\mu$ in a wide range of 
impact parameters. For example, a very sensitive $17 m$ diameter telescope, 
comprising a mirror area of $\sim 240 m^{2}$, would trigger on $\mu$ 
with energy $\geq 30 
GeV$ anywhere up to the so-called hump 
in the lateral distribution of Cherenkov light
at $\sim 120 m$ (see Fig.1).
The shape of a $\mu$ image depends on its energy and impact parameter and 
on the incident angle (for details see \cite{vacanti}). 
For any selected azimuth direction and for a short height interval a 
relativistic $\mu$ emits a parallel beam of light along the Cherenkov 
light cone opening angle $\theta_{Cher.}$. For simplicity, let us consider 
a $\mu$ that 
hits the telescope on-axis. Light from the $\mu$ from the given azimuth will 
be focused by the telescope optic into a spot that is 
at an angular distance of 
$\theta_{Cher.}$  
from the imaging camera center. Thus the emitted light 
from all the azimuth angles of the cone will be focused into a ring-shaped 
image of radius $\theta_{Cher.}$ in the focal plane. 
The thickness of the ring is 
largely due to variations in the Cherenkov light emission angle along 
the $\mu$ track. Close to the Cherenkov light emission threshold (that is, 
for example, $5 GeV$ for $\mu$ at $2200m$ a.s.l.) a $\mu$ 
can produce thick 
rings of smaller radii than the $\theta_{Cher.}$.       
A $\mu$ that hits anywhere in the reflector area of a telescope will produce 
a circular image of varying charge density along the circle and 
only in the case of an on-axis hit will it produce a "constant" 
charge density along the ring. The maximum height for a $\mu$ above the 
reflector area in which it still can produce a ring image is defined by 
the reflector's diameter and by the Cherenkov angle for the location 
height. For example, a $17m$ diameter IACT can see the light from 
on-axis $\mu$ below height of $\sim 400 m$. With the increase of the impact 
parameter up to $\sim 30 m$, a $\mu$ will produce image shapes that 
one can still recognize as an arc (see Fig.2). 
Beyond that, until the impact 
parameter of $\sim (60-70) m$, the arc image shape will shrink in 
length, essentially imitating short straight lines. The scarce photon 
statistics of the image and the usually relatively coarse pixel size 
of the imaging cameras make it even more difficult to reveal any hint of 
slight curvature in a short quasi-straight line. The images of $\mu$ shrink 
to essentially small spots (of high-charge concentration) for impact 
parameters $\geq 80 m$. The above-mentioned quasi-straight line images 
produced by $\mu$ can easily mimic the expected image shapes of $\gamma$. 
One can 
estimate that for the impact parameters range $\sim (30-60) m$, a telescope 
of $240 m^{2}$ reflector area will collect $\sim 1500$ photons 
from a $\mu$ (see Fig.1). 
Assuming an average photon to photo electron ($ph.e.$) conversion efficiency 
of $\sim 10 \%$ this will correspond to a $\sim 150 ph.e.$ signal 
for those $\mu$ events. 
That intensity is comparable to the image intensity produced 
by $\gamma$ showers 
near the threshold energy. 
This is possible because 
one can measure as many photons from a single long-flying $\mu$ in the 
field of view of a telescope as from many $e^{+}e^{-}$ of short 
track lengths together in an air shower.
In addition, because of the isotropic flux, 
some of those $\mu$ events will have the expected orientation of 
the $\gamma$ images 
and thus can pass the signal selection criteria. Those misclassified 
events can be numerous and thus can deteriorate the sensitivity of a 
telescope to genuine $\gamma$. 
Also stereoscopic telescope configurations, providing high photon 
sensitivity, will trigger on $\mu$ in a wide range of impact 
parameters.
A typical $\mu$ will produce 
arcs of different 
lengths in different telescopes, so one can tag it.
Frequently a stereoscopic telescope system will
measure $\mu$ events seen only by two telescopes (here we assume
a trigger configuration of "{\it any $\geq 2$ telescopes out of N}" 
where N is the total number of telescopes). Part of these $\mu$,
especially those coming from large impact parameters,
can mimic $\gamma$. Of course, compared to a single telescope
the rate of such events will be much lower.  
If the photon sensitivity 
of stereo telescopes, set into hardware coincidence, is not very high 
then only one of the telescopes, the closest one to the $\mu$ impact point 
can produce trigger. Thus such a system will strongly suppress that annoying 
background but only at the expense of relatively high energy threshold.
There is a first order correlation between the reflector diameter and
the track length of a particle seen in Cherenkov light.
At large impact distances a $10 m$ diameter dish collects light
from approximately one radiation length of the track of an
ultra-relativistic particle.
It is interesting to note that $\mu$ images from large impact parameters 
become elongated in the axial direction (see the image shape 
for $120 m$ impact distance on Fig.2). This reflects the variation of 
the refraction index and, correspondingly, of the Cherenkov light emission 
angle in the air when a $\mu$ traverses long distance in height,  
remaining in the telescope's field of view. 
In this article we want to report on ultra-fast timing features of $\mu$ 
events that can open a new dimension in Cherenkov technique allowing 
one to tag and to reject them.
As will be shown in the Monte Carlo (MC) section the characteristic 
time spreads are typically in the range of $(100-200) ps$ for 
$\mu$ images, 
$(2-3) ns$ for $\gamma$ and $(3-5) ns$ (with significant tails) 
for hadrons. 
In the following we will present detailed MC studies, discuss 
the possible performance and outline how a best-possible system 
for $\mu$ suppression should look. 
\section{Triggers from muons: a closer look}
\subsection{Expected ultra-short pulse in time}
The arrival time distribution of Cherenkov photons from extensive air 
showers has been extensively studied (see, for example, \cite{chitnis}, 
\cite{cabot}). One may 
assume that measured photon arrival time structures from an air shower 
of a given impact parameter and of a given incident angle are due to  
\begin{itemize}
\item
the shower's transverse and longitudinal size seen by the telescope, 
the longitudinal size being the dominant effect 
\item
shower particle velocity differences from the speed of light 
along the shower's longitudinal development 
\item
energy-dependent multiple-scattering angle of shower particles 
along the shower height
\item
angular deviation of produced particles from the primary direction  
\end{itemize}
The elastic scattering effects in the earth's magnetic field 
may also contribute into the time structure. 
Unlike air showers a single $\mu$ has no transverse size.
Also the last 
point in the above list cannot contribute into the time
structure of light pulse from a $\mu$. 
The light from a $\mu$ has a time 
structure only because of the other criteria above. 
Calculations show that light emitted at a height of
$35 km$ will arrive $\sim 6 ns$ later than the 
light emitted immediately above the ground level
due to the higher speed of a relativistic $\mu$ compared to
that of light in the atmosphere. 
On the other 
hand, a telescope, because of it's limited field of view can 
only observe a small part of the track of a $\mu$ (as outlined in 
the previous chapter) that will, correspondingly, produce a very 
short flash. 
	To quantify the effect of the other influences we have 
performed Monte Carlo simulations. 
\subsection{Monte Carlo simulations}
We used the version 6.023 of the CORSIKA code in our simulations \cite{knapp}. 
Our statistics is based on $7.5 \times 10^{5}$ simulated proton showers in the 
energy range $(0.04-30) TeV$ following a power-law of index -2.75. The 
impact parameter range for the protons was up to $300 m$. They were 
simulated from a zenith angle of $20^{\circ}$ and within the cone opening 
angle $2.5^{\circ}$. All $\mu$ produced were tracked. 
Along with protons 35000 $\gamma$ showers were simulated in the energy 
range $(0.01 - 30) TeV$ following a power-law of index -2.6, impact parameter 
range up to $300 m$ and the zenith angle of $20^{\circ}$. Also $4 \times 
10^{5}$ 
$\mu$ were 
simulated in the energy range $(10-100) GeV$, following a power-law of 
index -2.69 and impact parameter range up to $200 m$, within $1.2^{\circ}$ 
around the 
observation direction. These $\mu$ were injected into the atmosphere at 
$100 g/cm^{2}$ (corresponding to height in the atmosphere of $\sim 17 km$). 
Depending on the problem under study we used different samples of 
the simulated events. In addition, we simulated $ 5 \times 10^{5}$ proton 
showers of fixed energies 50, 100, 300 and 500 $GeV$ impinging from 
the zenith angle $0^{\circ}$. This sample was used to produce Fig.3a in which 
we show the average number of produced $\mu$ versus the altitude in the 
atmosphere (for every $1km$ height interval) and Fig.3b in which we show 
the same but only for $\mu$ that are above the Cherenkov light
production threshold for the given height.
The areas under the corresponding 
histograms on Fig.3 are the average number of $\mu$ produced in a shower 
of a given energy. One can see that the maxima of $\mu$ production for 
different primary energies are around the height range of $10-12 km$.
The 50, 100, 300 and 500 $GeV$ proton samples were 
used to produce Fig.4 which shows the probability of angular 
deviation of $\mu$ from the primary direction. The angle is measured in 
the last part of the $\mu$ trajectory just before it hits the observation 
level. In Fig.4a one can see the angular deviation for all $\mu$ and in 
Fig.4b only for those that are above the light production threshold 
(the area under the distributions are normalized to 1). While at high 
energies the angular distribution can be characterized by a 
$\sim 1^{\circ}$, it 
is wider (with a tail extending up to $(4-6)^{\circ}$) for 
lower energies (see 
the curve produced by $50 GeV$ protons).
The qualitative picture is clear: at low energies the secondary particles 
are at wider angles to the primary direction (read: the probability is 
high that a $\mu$ can appear as a "single" particle in the limited field 
of view of the 
imaging 
camera). 
By switching the earth's 
magnetic field on and off in Monte Carlo simulations, we found that it 
does not influence the width and the shape of the angular deviation of 
the $\mu$. Also, the estimates of the multiple scattering effect 
of $\mu$ provide 
lower value of the deviation angle from the primary direction than the 
results of the simulations show. The reason for this difference can be 
due to the known non-Gaussian larger tails in the real scattering process.   
Fig.5 shows the characteristic photon arrival time profiles for $\mu$, 
$\gamma$ and 
hadrons at the observation level of $2200 m$ a.s.l.. One can see that the 
time structure of light flashes from $\mu$ is $(100-200) ps$ wide. 
There is a 
small tail in the distribution where $\mu$ produce somewhat longer pulses 
but, as one can see, those happen rarely. Typically, the $\gamma$ 
flashes 
at low energies are $(2-2.5) ns$ wide while the majority of protons 
produce 
flashes of $(2-6) ns$ wide with a tail extending beyond $10 ns$. Also, one can 
see in Fig.5 that the photon arrival time distribution from $\gamma$ 
showers 
has nearly Gaussian shape, while the same from protons is significantly 
asymmetric. 
\section{Muon tagging by a telescope}
The measured signal from a telescope is a convolution of the incident 
light time profile with the response functions of a) the reflector, b) 
the light sensors and c) the readout system. When identifying $\mu$, 
because of their ultra-fast flashes 
(see Fig.5), one would search for pulses close to the
instrumental response to instantaneous charge injection. Anything larger
than that will be 
due to other reason than the detection of a $\mu$. In the ideal case, the 
instrumental response function must be ultra-fast in order not to 
smear the time signature of $\mu$ and to provide discrimination power. 
Below we list the main parameters that determine the instrumental 
time response of a telescope:
\begin{itemize}
\item
design of the reflector
\item
speed of the focal plane light sensors
\item
speed of the data acquisition system 
\end{itemize}
Let us discuss the above listed factors.\\ 
{\it Design of the reflector}: it is well known that a parabolic reflector 
has no time dispersion 
at the focal plane for a parallel beam of light arriving on-axis.
For an 
off-axis parallel beam of light
inclined to $1^{\circ}$, 
for a parabola of $F/1$ optics, 
one will obtain a time dispersion of slightly less than $140 ps$. 
Calculations show that for larger angles, up to a few degrees, that 
value increases only marginally. Usually, the field of view of 
relatively fast design parabolic reflectors ($F/D \sim 1$), which 
can provide optical 
resolutions of $\sim (0.05-0.1)^{\circ}$ acceptable for air Cherenkov
telescopes, is limited 
to $(1-1.5)^{\circ}$ in radius, so the 
time dilution caused by the above mentioned effect will be 
small. One may conclude that a reflector of parabolic design is 
appropriate for ultra-fast timing purposes (see, for example, 
\cite{mariotti}).
There can be other factors which deteriorate the timing profile but 
usually all of these will have minor contributions. There can be 
contributions because of the 
\begin{itemize}
\item
tessellated design of the reflector; the shape of individual 
mirrors deviate at the edges from the perfect parabola. This effect 
shall be below $100ps$, even in the case of fast optics and of relatively 
large mirror tiles of $\sim 1 m$ size.
\item
differences in the fixation of the mirror tiles on the reflector 
that could result in deviation from the parabolic shape. One may assume 
that this effect can be relatively easy controlled and that the 
deviations can be kept below $2 cm$. That will result in time smearing 
of $\leq 120 ps$.
\item
use of light concentrators in front of the sensors in the 
camera to minimize the light losses. Light can hit the 
sensors directly or via single or double reflections from the concentrator. 
Path differences of $\leq (2-3) cm$ will result in time smearing of 
$\leq 100 ps$.      
\end{itemize}
Several other reflector designs exist that, unfortunately, always 
expand the time profile of input light. The commonly used Davies-Cotton 
design widens the input pulse, even for the on-axis light flash, to for 
example, $(0.3 ns/m) \times F(m)$ for $F/D = 1.2$.
\\
{\it Ultra-fast light sensors}: until now the classic photo-multiplier tubes 
(PMT) have been the only light sensors used in the focal plane imaging cameras 
of Cherenkov telescopes. Usually their response is in the range of 
$(2-3) ns$ 
but very fast tubes also exist. For example, in the MAGIC telescope 
project, 6-dynode, hemi-spherical bialkali $1"$ PMTs from Electron Tubes 
(ET9116A) are used. They provide rise  and fall times of 0.6ns and 0.7ns, 
respectively, and a pulse width of $(1.0-1.2) ns$. The measured transit time 
spread (TTS) is $\leq 300ps$. So a $\delta$-function-like 
light flash at the input of 
these PMTs will produce an ultra-short pulse with the above-mentioned 
parameters. 
\\
{\it Ultra-fast readout system}: in order to provide any 
meaningful pulse shape 
reconstruction, assuming a Gaussian-shaped input pulse profile, one needs 
at least $3-4$ sampling points. By using an ultra-fast readout system 
of $\geq 2 Gsample/s$ one can reconstruct the very fast pulse shapes from, 
for example, above mentioned ultra-fast PMTs (see, for example, 
\cite{multiplexer}). 
\section{Discussion: $\mu$ suppression by ultra-fast timing}
From the discussion above one may conclude that it is currently possible 
to design a telescope that can produce ultra-short signal pulses of 
$\sim 1.5 ns$,
as a response to $\delta$-function 
like light flashes 
(as for example, from $\mu$). 
The $\mu$ flashes will generate output pulses that should be only marginally 
wider than the instrumental response function of a telescope. This can 
be used as a criterion to distinguish them from $\gamma$ and 
proton air showers. 
Although Cherenkov flashes produced by $\gamma$ showers are significantly 
shorter and have smoother time profile compared to those produced by 
hadrons (see below), still they have $\sim (2-3) ns$ time structure that is 
significantly wider than that of $\mu$.  
We have also calculated the root mean squared $(r.m.s.)$ value of the 
arrival time distribution for all $\gamma$, proton and $\mu$ 
events measured by 
a $17 m$ telescope with an ultra-fast response function. In addition to 
the above-mentioned parameters for the PMT, we simulated output pulses 
with realistic transit time spread (TTS) of $\sigma = 300 ps$. 
We have used a 
trigger condition {\it any 4 next neighbor pixels above a given 
threshold} in our simulations. The corresponding distributions are 
shown on Fig.6. One can see a good separation of the $\mu$ events 
from $\gamma$. 
In fact, there are no $\gamma$ events with $r.m.s. \leq 0.7 ns$ or with 
$r.m.s. \geq 3.5 ns$. All events with $r.m.s.$ time spread $\leq 0.7 ns$ 
are $\mu$.
It is also interesting to study the dependence of the $r.m.s.$ time width 
on the image {\it SIZE} (measured as the sum of detected $ph.e.$). The 
corresponding distributions are shown in Fig.7a. The $r.m.s.$ 
distributions of the $\gamma$ and of the $\mu$ overlap in small 
{\it SIZE} region. 
When applying a simple set of loose supercuts based on the {\it length, 
width, distance} and the {\it alpha}
parameters
for $\gamma$/hadron separation, 
the situation is almost the same (see Fig.7b). 
In contrast,
the
new cut on the
$r.m.s.$ time spread can strongly 
suppress the unwanted $\mu$
background.
\subsection{Gamma telescopes based on solar arrays}
Note please that the above mentioned is equally true for imaging 
air Cherenkov telescopes and for solar array type $\gamma$ telescopes. 
The latter 
telescopes 
(for a recent review see \cite{smith})
shall also experience $\mu$ background. On the other hand, as 
a rule, those instruments have an ultra-fast response and a readout that 
could help to tag $\mu$. Application of the proposed new technique can 
improve the sensitivity of the solar array-type $\gamma$ telescopes.
\subsection{Stereoscopic telescope systems and wide field of view telescopes}
The imaging stereoscopic telescope systems, providing high photon
sensitivity and low threshold, will effectively trigger on $\mu$.
Due to the multiple views and produced different arc
lengths and shapes in different telescopes, as a rule one can tag the
$\mu$. Still some part of $\mu$, especially those from large impact 
parameters, can survive the image cuts and mimic $\gamma$. \\
A wide field of view telescope could offer another remedy
to tag the $\mu$: frequently one can detect also their parent
showers (or particles) that will help the identification.
It seems that also in this case part of the $\mu$ could be 
misclassified,
especially those detected near the camera edges.\\
An ultra-fast telescope design will help both in the case of the
stereo and wide field of view telescopes 
allowing one to effectively tag the 
misclassified $\mu$.\\

\section{Conclusions}
We have shown that light pulses from single, long-flying relativistic 
charged particles 
have an ultra-narrow time structure. The differences in pulses generated 
by the $\mu$, $\gamma$ and  hadron events are large enough so that one can 
efficiently tag and suppress the $\mu$ events. 
The new method suggested here 
opens a new dimension for the solution of the so-called "$\mu$-wall" 
problem: ultra-fast telescopes, both of imaging and 
non-imaging types, including the solar array types, can provide 
very efficient $\mu$ suppression.\\ 
The new method can directly increase the sensitivity  
of the ground-based $\gamma$ ray telescopes.  
\section*{Acknowledgements}
We are grateful to Dr. K. Mase for providing some of the simulation 
results. D. S. wants to acknowledge the Polish KBN grant No. 1P03D01028.
\newpage

\begin{figure}[h!]
\begin{center}
\includegraphics[totalheight=7cm]{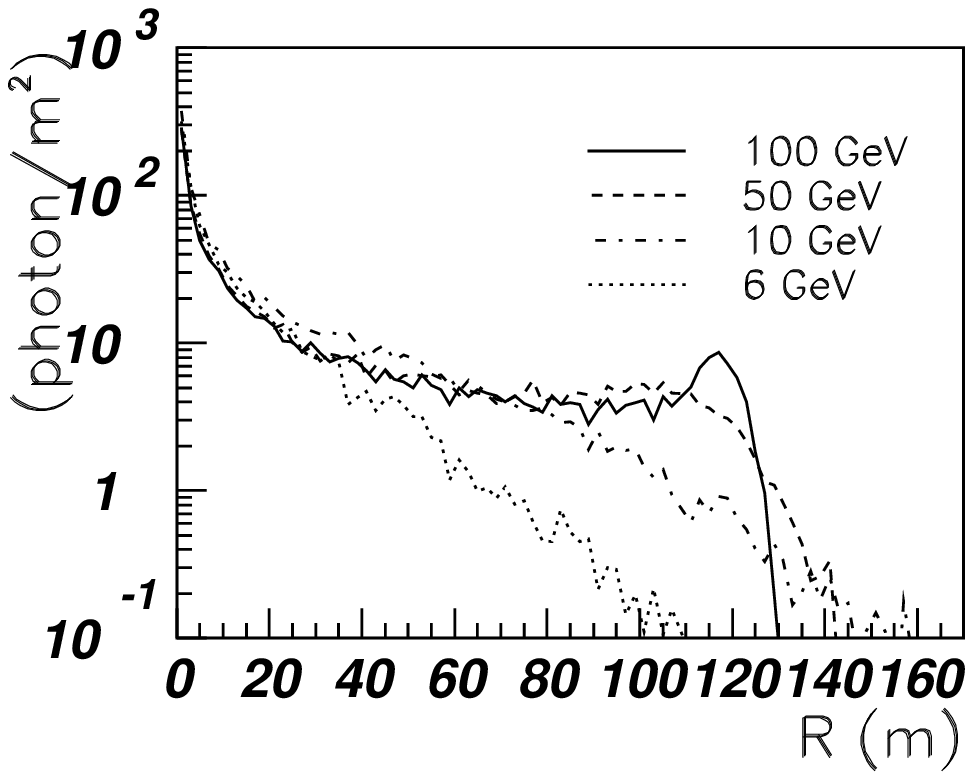}
\end{center}
\caption{\it Simulated lateral distribution of Cherenkov light density on 
the 2200 m a.s.l. from a 6, 10, 30 and 100 GeV $\mu$, injected at 
a height of 17 km in the atmosphere.}
\end{figure}

\begin{figure}[h!]
\begin{center}
\includegraphics[totalheight=7cm]{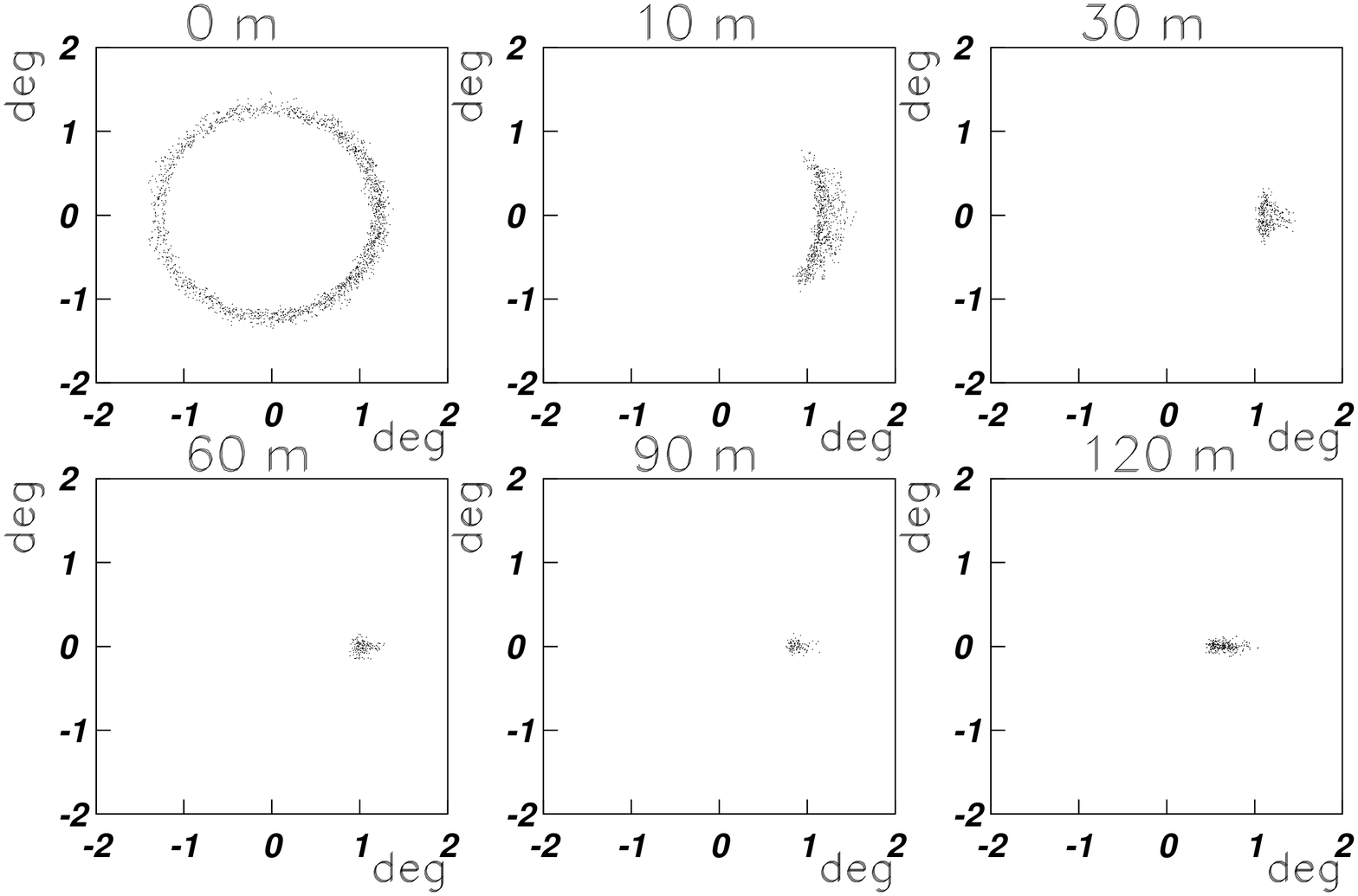}
\end{center}
\caption{\it Induced by $\mu$ arc-shape images in the focal plane 
of a telescope as function of impact parameter. The images 
include optical aberrations and are simulated for a 17m 
diameter telescope.}
\end{figure}
\begin{figure}[h!]
\begin{center}
\includegraphics[totalheight=7cm]{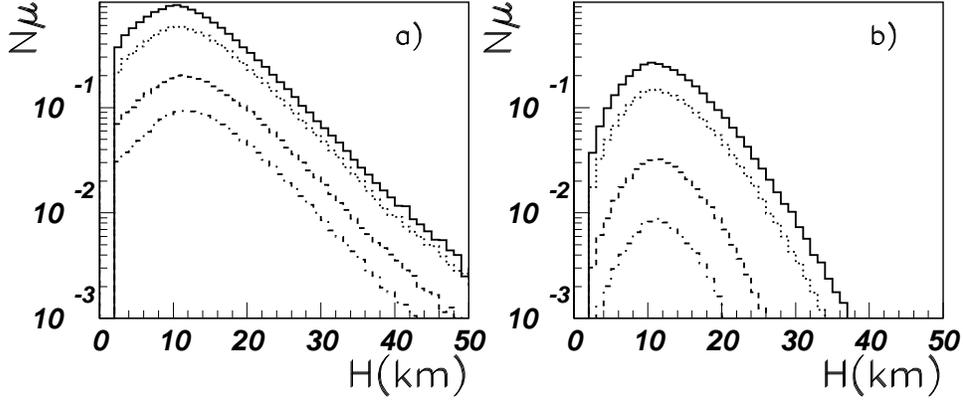}
\end{center}
\caption{\it Number of $\mu$ in air showers versus the production 
height in the atmosphere. The distributions for proton 
energies of 50, 100, 300 and 500 GeV are shown, 
from the bottom upwards:
a) all $\mu$; b) only $\mu$ that are above the Cherenkov threshold
for the given height.}
\end{figure}
\begin{figure}[h!]
\begin{center}
\includegraphics[totalheight=7cm]{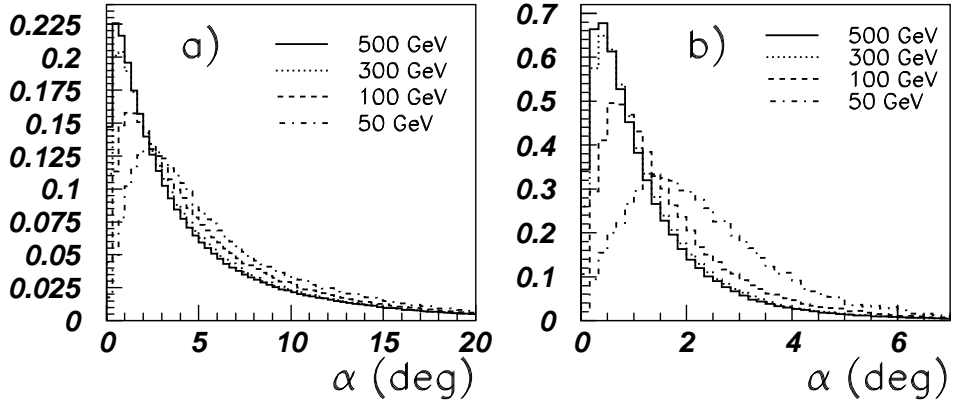}
\end{center}
\caption{\it Probability of angular deviation of $\mu$ (in fact, the 
angle of $\mu$ intersection with the observation level) from the 
primary proton direction, shown for primary energies of 50, 100, 
300 and 500 GeV: a) all $\mu$; b) only $\mu$ above the 
Cherenkov threshold at corresponding heights. The areas under 
the distributions are normalized to 1.}
\end{figure}
\begin{figure}[h!]
\begin{center}
\includegraphics[totalheight=7cm]{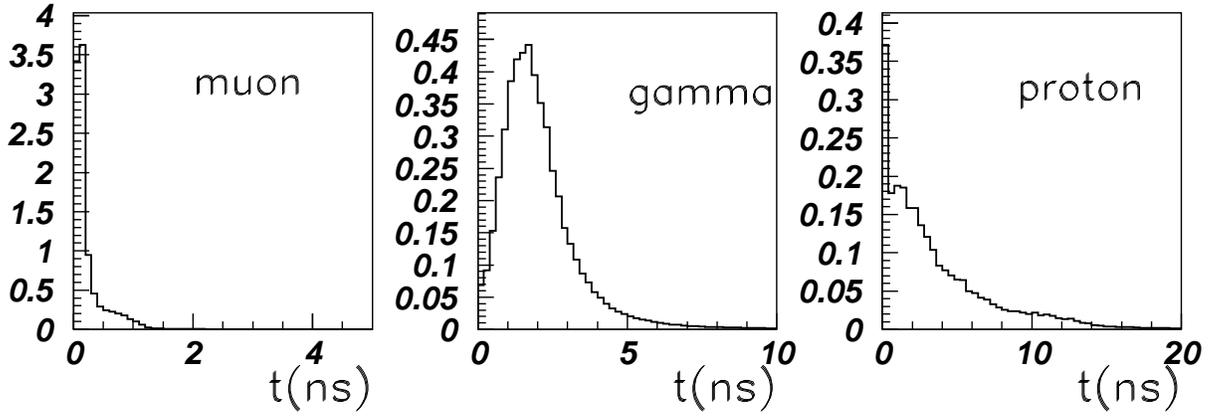}
\end{center}
\caption{\it Photon arrival time distributions for $\mu$, $\gamma$ and  
protons. The distribution for $\mu$ is essentially below $(100-200) ps$. 
Photons from a $\gamma$ shower (from a weighted spectrum) arrive within  
$(2-2.5) ns$ while photons from a proton shower arrive essentially in 
$(2-6) ns$ with a tail extending beyond $10 ns$.}
\end{figure}
\begin{figure}[h!]
\begin{center}
\includegraphics[totalheight=7cm]{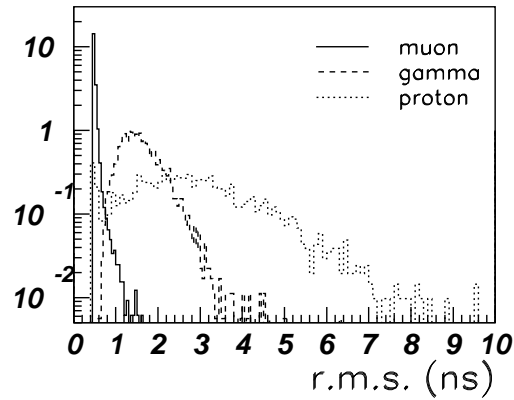}
\end{center}
\caption{\it Distributions of the $r.m.s.$ value of the arrival time  
distribution for all $\gamma$, proton and $\mu$ events measured 
by a 17m ultra-fast telescope. The events with $r.m.s. 
\leq 0.7 ns$ are of $\mu$ origin.}
\end{figure}
\begin{figure}[h!]
\begin{center}
\includegraphics[totalheight=7cm]{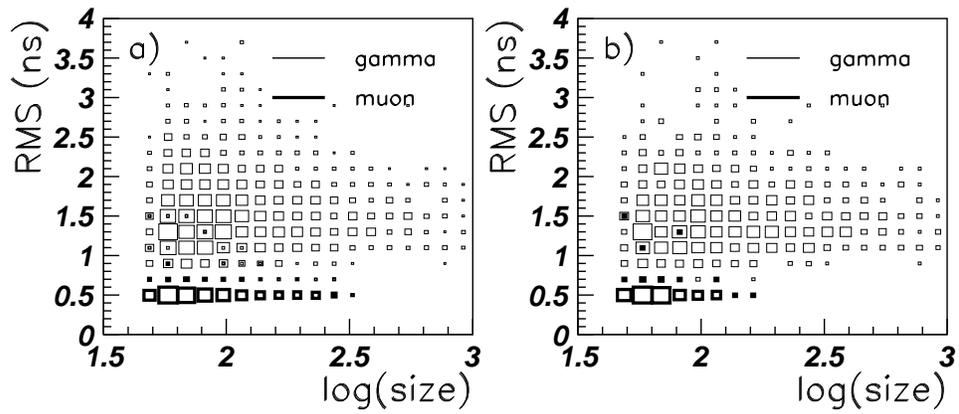}
\end{center}
\caption{\it Dependence of the $r.m.s.$ time spread of the $\gamma$  
and $\mu$ images on the parameter {\it SIZE} (sum of registered ph.e.). 
b) $r.m.s.$ dependence on {\it SIZE} after application of simple set 
of supercuts.}
\end{figure}

\end{document}